\documentclass[aps,prd,twocolumn,nofootinbib]{revtex4-2}

\newcommand\ForInternalReference[1]{}
\newcommand\SkipForEarlyCirculation[1]{}

\newcommand\SkipPP[1]{}
\usepackage{verbatim}
\usepackage{graphicx}
\usepackage{dcolumn}
\usepackage{bm}
\usepackage{color}
\usepackage{multirow}
\usepackage{xspace}
\usepackage{url}
\usepackage{amsmath}
\usepackage{xcolor}
\usepackage{float}
\usepackage{listings}
\usepackage{multirow}
\usepackage{amssymb}
\usepackage{color}
\usepackage{times}
\usepackage{tikz}
\usepackage{makecell}
\usetikzlibrary{shapes.geometric, arrows, bayesnet}
\usepackage[colorlinks=true,linkcolor=blue,citecolor=blue]{hyperref}
\newcommand\optional[1]{}
\newcommand\revision[1]{{\color{black}#1}}

\tikzstyle{startstop} = [circle, rounded corners, minimum width=1cm, minimum height=1cm,text centered, draw=black, fill=red!30]
\tikzstyle{io} = [trapezium, trapezium left angle=70, trapezium right angle=110, minimum width=2cm, minimum height=1cm, text centered, draw=black, fill=blue!30]
\tikzstyle{process} = [rectangle, minimum width=2cm, minimum height=1cm, text centered, draw=black, fill=orange!30]
\tikzstyle{decision} = [diamond, minimum width=2cm, minimum height=1cm, text centered, draw=black, fill=green!30]
\tikzstyle{arrow} = [thick,->,>=stealth]
\definecolor{amber}{rgb}{1.0, 0.75, 0.0}
\definecolor{orange}{rgb}{1.0, 0.5, 0.0}
\definecolor{amaranth}{rgb}{0.9, 0.17, 0.31}

\graphicspath{{./figures/}}
\newcommand{\mc}{{\cal M}}
\newcommand{\Ms}{M_{\odot}}

\def\ltsima{$\; \buildrel < \over \sim \;$}
\def\simlt{\lower.5ex\hbox{\ltsima}}
\def\gtsima{$\; \buildrel > \over \sim \;$}
\def\simgt{\lower.5ex\hbox{\gtsima}}

\newcommand\gwk{\textsc{GWKokab}\xspace}
\newcommand\flowMC{\textsc{flowMC}\xspace}

\newcommand{\srate}{\boldsymbol{\mathcal{R}}}
\newcommand{\pvec}{\Lambda} % population parameter vector
\newcommand{\svec}{\lambda} % source properties, m1,m2,a1,a2
\newcommand{\zvec}{\kappa} % redshift evolution parameter
\newcommand{\comp}{\boldsymbol{\rho}}
\newcommand{\svecz}{\boldsymbol{\lambda}} % source properties, m1,m2,a1,a2, z ..
\newcommand{\pvecz}{\boldsymbol{\Lambda}}
\newcommand{\data}{\mathcal{D}}

\def\RIT{Center for Computational Relativity and Gravitation, Rochester Institute of Technology, Rochester, New York 14623, USA}
\def\Habib{Habib University, Karachi, Pakistan}

\begin{document}

\renewcommand{\arraystretch}{1.5}
\title {An Implementation to Identify the Properties of Multiple Population of Gravitational Wave Sources}

\author{M. Qazalbash}
%\email{meesumqazalbash@gmail.com}
\affiliation{\Habib}

\author{M. Zeeshan}
\email{m.zeeshan5885@gmail.com}
\affiliation{\RIT}

\author{R. O'Shaughnessy}
%\email{richardoshaughnessy.rossma@gmail.com}
\affiliation{\RIT}

%%%%%%%%%%%%%%%%%%%%%%%%%%%%%%
\begin{abstract}

	The rapidly increasing sensitivity of gravitational wave detectors is enabling the detection of a growing number of compact binary mergers. These events are crucial for understanding the population properties of compact binaries. However, many previous studies rely on computationally expensive inference frameworks, limiting their scalability.
	In this work, we present \gwk, a \textsc{JAX}-based framework that enables modular model building with independent rate for each subpopulation such as binary black hole, binary neutron star, and neutron star black hole binary. It provides accelerated inference using the normalizing flow based sampler called \flowMC and is also compatible with \textsc{NumPyro} samplers.
	To validate our framework, we generated two synthetic populations, one comprising spinning eccentric binaries and the other circular binaries using a multi-source model. We then recovered their injected parameters at significantly reduced computational cost and demonstrated that eccentricity distribution can be recovered even in spinning eccentric populations. We also reproduced results from two prior studies: one on non-spinning eccentric populations, and another on the binary black hole mass distribution using the \revision{fourth Gravitational Wave Transient Catalog (GWTC-4)}.
	We anticipate that \gwk will not only reduce computational costs but also enable more detailed multi population analyses such as their mass, spin, eccentricity, and redshift distributions, offering deeper insights into compact binary formation and evolution.

\end{abstract}

\maketitle

%\linenumbers

\section{Introduction}
\label{sec:intro}

%\textbf{L.Review}

The first direct detection of gravitational wave (GW) by the Laser Interferometer Gravitational-Wave Observatory (LIGO) \cite{2016PhRvL.116f1102A} opened a new observational window to the universe and the detectors \cite{2018LRR....21....3A, 2015CQGra..32g4001L,2015CQGra..32b4001A,2021PTEP.2021eA101A} are enabling us to probe phenomena inaccessible through electromagnetic observations \cite{2020ApJ...896L..44A, 2009LRR....12....2S}. GWs are emitted during the inspiral and merger of compact objects such as binary black hole (BBH), binary neutron start (BNS), and neutron star black hole (NSBH) systems carrying rich information about their astrophysical origins \cite{2009LRR....12....2S, 1977ASIB...27....1T}. Since the initial detection, the number of observed GW events has grown rapidly \cite{2019PhRvX...9c1040A,2021PhRvX..11b1053A,2024PhRvD.109b2001A,2021ApJ...922...76N,2022PhRvD.106d3009O,2023ApJ...946...59N, 2025arXiv250818079T,2025arXiv250818081T}, a trend expected to continue as detector sensitivities improve \cite{2015CQGra..32g4001L,2015CQGra..32b4001A,2019NatAs...3...35K,2021PTEP.2021eA101A,2025arXiv250818081T}. This expanding catalog enables population studies that yield insights into merger rates, mass, spin redshift, and orbital eccentricities distributions \cite{2023PhRvX..13a1048A,2022PhRvD.106j3019T,2017Natur.548..426F,2018ApJ...868..140T,2019PhRvD.100d3012W,2025arXiv250818083T}. These analyses not only improve our understanding of compact binary formation and evolution \cite{2016Natur.534..512B,2022ApJ...940..171R,2020ApJ...903L...5R,2021ApJ...921L..43Z} but also offer stringent tests of general relativity \cite{2023PhRvD.108b4043M,2023PhRvD.107d4020M,2021PhRvD.103l2002A,2016PhRvL.116v1101A,2019PhRvL.123l1101I}. Recent observational evidence suggests that compact binary mergers may originate from multiple formation channels \cite{2016PhRvL.116b9901R,2017PhRvD..95l4046G,2016Natur.534..512B}, such as isolated binary evolution in the field, dynamical interactions in dense stellar environments, or hierarchical mergers. Properly modeling these diverse formation channels requires the ability to construct and analyze mixture models with independent rate and parameter distributions \cite{2021ApJ...910..152Z,2022ApJ...936L..18G,2022ApJ...941L..39W,2021PhRvD.104h3010R}, a capability that is limited in many existing frameworks.

%\textbf{Previous codes}

Several computational frameworks have been developed to infer the population properties of compact binaries, including parametric and non-parametric approaches. Notable examples include \textsc{PopModels} \cite{git_popmodels}, \textsc{GWPopulation} \cite{2019PhRvD.100d3030T,2019ApJS..241...27A,2021zndo...5654673T}, \textsc{GWInferno} \cite{2023ApJ...946...16E,git_gwinferno}, \textsc{Sodapop} \cite{git_sodapop}, \textsc{ICAROGW} \cite{2024A&A...682A.167M,git_icarogw}, \textsc{GWMockCat} \cite{2020ApJ...891L..31F,git_gwmockcats}, and variational inference approach in \textsc{gwax} \cite{git_gwax,2025PhRvD.111l3049M}. In addition, deep learning and machine learning approaches have emerged to infer population properties from GW catalogs \cite{2020PhRvD.101l3005W,2024PhRvD.109f4056L,2020arXiv201201317T,2021PhRvD.104h3531G,2022PhRvD.106j3013M,2022arXiv221109008R,2025ApJ...991...17R,2021CQGra..38o5007T}. While parametric methods have improved in flexibility and scalability, they often remain computationally intensive especially in modeling subpopulations or mixture models. Among these, \textsc{PopModels} \cite{git_popmodels} is one of the few tools capable of characterizing subpopulations with their independent rates and spins, but it lacks the computational efficiency required for large-scale inference and very slow on the growing dataset. Besides slower speed \textsc{PopModels} does not offer the independent redshift and eccentricity distributions which is important to study the formation and evolution of the binaries.

%\textbf{Use case} and injection recovery
\gwk~\cite{git_gwkokab} addresses these limitations by integrating modern computational tools and statistical techniques along with user-friendly flexibility to construct complex models from simple components such as multi-source. Unlike traditional frameworks that rely on slower sampling methods and rigid model structures, \gwk leverages hardware acceleration via \textsc{JAX} \cite{jax2018github} and normalizing flows based sampling through \flowMC \cite{git_flowMC,2023JOSS....8.5021W,2022PNAS..11909420G}. In addition to \flowMC, \gwk is also compatible with \textsc{NumPyro} \cite{phan2019composable,bingham2019pyro} which requires lesser GPU memory on heavy datasets and makes it suitable for multi-source models. Furthermore, \gwk also allows to study independent eccentricity and redshift distributions of subpopulations unlike \textsc{PopModels} which only allows for independent mass and spin distributions.
This combination enables scalable, high-dimensional inference with improved sampling efficiency, even for complex and multi-modal parameter spaces.

\gwk also supports mock catalog generation with independent rates for potential science studies. One can build a model, generate injections based on provided sensitivity and then may add pre-defined errors \cite{2019MNRAS.486.1086M} in each parameter of the event or they can also use \textsc{RIFT} \cite{2022arXiv221007912W} to add realistic error using desired choice of waveform. It also allows to build models to generate eccentric events and make inference on eccentric population models. We tried to keep the framework adaptable and efficient for future studies, allowing future user or developer to add new models.
To validate the performance of \gwk, we generated a synthetic population of spinning eccentric BBHs and successfully recovered the injected parameters. Additionally, we created a multi-source population of BBHs, BNSs, and NSBHs modeled with independent rates as described in \cite{2023PhRvX..13a1048A} and recovered both the population parameters and their respective rates.

%\textbf{Reproduced results }

To show the correctness, we also reproduced two previously published results. First, we replicated the key findings of the eccentricity matters study \cite{2024PhRvD.110f3009Z} at 98\% lower computational cost, using the same model, priors, and volume-time sensitivity injections that were trained on a \revision{neural net defined in Section \ref{subsubsection:neural_net_pdet}}. Second, we replicated the BBH population study based on \revision{153 BBHs from GWTC-4 \cite{2025arXiv250818083T}}, using the same model and semi-analytical sensitivity injections. In this paper, we demonstrate the capabilities of \gwk through several scientific use cases. These include recovery of synthetic spinning eccentric BBH populations, inference on multi-source models with independent rates and spins, and reproduction of key results from previously published population studies. These examples highlight the computational efficiency and flexibility of \gwk for large-scale gravitational-wave population inference.

%\textbf{outline}
This paper is organized as follows: Section \ref{sec:methods} explains the methods used to develop \gwk, we explained hierarchical Bayesian inference starting from individual event to population inference. In Section \ref{sec:validation_analyses}, we present validation studies with real and synthetic populations. Finally, Section \ref{sec:conclude} summarizes the key findings and outlines directions for future work. The necessary technical details are provided in Appendix \ref{sec:appendix}.

\section{Methods}
\label{sec:methods}
We used the following methods in \gwk to infer the population properties of compact binary mergers using gravitational wave data. The framework is designed to be modular, allowing users to build complex population models from simple components.

\subsection{Population Model Construction}
\label{sec:models}

A population model describes the distributions of intrinsic properties of
compact binary mergers, such as mass, spin, and eccentricity.
In this subsection, we follow notation established in \cite{2025PhRvD.111f3043H}, using conventions which reduce to the
notation adopted in \cite{2019PhRvD.100d3012W}.
In our framework, source parameters characterized by \(\svec\) excluding redshift.  While nominally $\svecz$ includes all intrinsic and
extrinsic parameters, without loss of generality we suppress most extrinsic parameters with naturally geometric uniform priors such as source
orientation, polarization angle, sky position, and event time; the manifold of source parameters is assumed to have some
metric with determinant $g_\svec$.
%intrinsic source parameters (mass, spin, eccentricity) are presented by \(\svec\), and \(\svec\) excludes extrinsic variables such as redshift denoted by $z$.
The population parameters $(m_{\mathrm{min}}, m_{\mathrm{max}}, \alpha, \beta, \cdots, \mu, \sigma)$ are denoted by
\(\pvec\).
The models is fully characterized by its detector frame merger rate per unit population parameters: $dN/dt_d d\svec
	\equiv R(\svec |\pvec)$ with units
$({\rm yr}^{-1} \svec^{-1})$, the local merger rate density in detector frame of reference under population parameters $\pvec$.
%From this toIn this context, \(p(\svec|\pvec)\) is the parametric probability density of intrinsic source parameters $\svec$ conditioned on the population parameters 

The total detector frame merger rate per unit population parameters can be decomposed into contributions from multiple
populations.  Assuming we have \(M\) number of populations with different population parameters \(\pvec_i\) for \(i=1,2,\cdots,M\), then $R_i(\svec)$ for each population is defined as follows
\begin{align}
	\label{eq:local_detector_rate}
	R_i(\svec) = \frac{dN_i}{d\svec dt_d} = \frac{1}{T} \frac{dN_i}{d\svec}.
\end{align}
where $N_i$ is the number of events in the $i^{th}$ population.

The merger rate density $\srate_i(\svec)$ in source frame of reference is more relevant for astrophysical population models and is defined
as the number of events per unit comoving volume per unit time with units $(\rm Gpc^{-3} yr^{-1} \svec^{-1})$, and is given by
\begin{align}
	\label{eq:source_rate_uniform}
	\srate_i(\svec) = \frac{dN_i}{dV_c dt_s~d\svec} = R_i(\svec) \left((1+z)^{-1}\cdot \frac{dV_c}{dz}\right)^{-1}.
\end{align}
where  $t_s=t_d(1+z)^{-1}$ is the source frame time related to detector frame time $t_d$, and $dV_c/dz$ is the
differential comoving volume per unit redshift $z$, which is a function of redshift $z$, and accounts for the expansion
of the universe. Importantly, $\srate_i(\svec)$ is merger rate density over intrinsic parameter $\svec$, not a
density of the extrinsic parameter $z$.

In general, the models $R_i(\svec)$ or equivalently $\srate_i(\svec)$ depend on redshift. Rather than allow for
flexible or high-dimensional redshift evolution, for simplicity and following previous work \cite{2023PhRvX..13a1048A} we adopt a simple power law
model for redshift dependence given as,
\begin{align}
	\label{eq:source_rate_redshift}
	\srate_i(z) & = \srate_{0_i} \cdot (1+z)^{\zvec_i} \propto (1+z)^{\zvec_i}.
\end{align}
where $\srate_{0_i}$ is the local merger rate density at redshift $z=0$ in source-frame of reference and the exponent
$\zvec_i$ is the redshift evolution parameter for the $i^{th}$ population.   Setting $\zvec=0$ implies no redshift
evolution (constant comoving volume), while $\zvec>0$ implies an increasing merger rate with redshift, and $\zvec<0$
implies a decreasing merger rate with redshift.

% For notational convenience, we explicitly factor out $\zvec$ from the remaining population parameters $\pvec$. so a single component of the overall population has the following
% relationship between $R(\svec)$ and $\srate(\svec)$:
% \begin{align}
%     \label{eq:source_rate_evolving}
%     \srate_i(\svec|\zvec_i)  = R_i(\svec) (1+z)^{\zvec_i+1} \left(\frac{dV_c}{dz}\right)^{-1}.
% \end{align}

For each component, we can decompose $\srate_i$ into a normalization factor and a nominal source-frame probability density
$p(\svec|\pvec) $, simply from the ratio of $\srate_i$ to
its integral over all population parameters. %is the parametric probability density of intrinsic source parameters $\svec$ conditioned on the population parameters 
In practice, we perform our calculations on a finite interval $[0,z_{\mathrm{max}}]$ so that the integrals needed for this
decomposition remain finite. % and to keep the original expression consistent, we multiply this normalization factor with $R_i(\svec)$. 
If we define $\srate^*(\pvec) = \int \srate (\svec) d \svec$, then for any component, we have the decomposition
\begin{align}
	\label{eq:source_rate_decomp}
	\comp_i(\svec,z\mid\pvec_i,\zvec_i) = \srate^*_i(\pvec_i)  p_i(\svec|\pvec_i) (1+z)^{\zvec_i},
\end{align}
where $\srate^*$ has units $(\rm Gpc^{-3} yr^{-1})$. The term $p_i(\svec|\pvec_i)$ is the normalized probability
density of intrinsic source parameters conditioned on the population parameters $\pvec_i$ and is assumed to be
independent of redshift $z$. For redshift-independent models with $\zvec=0$, the prefactor
has a natural interpretation as the overall merger rate, and of course the total merger rate density in source frame of reference is then given by summing over all population rates, which can be expressed as
\begin{align}
	\label{eq:total_rate}
	\comp(\svecz\mid\pvecz) = \sum_{i=1}^{M} \comp_i(\svecz\mid\pvecz_i).
\end{align}

where bold face $\svecz$ characterize all intrinsic and extrinsic binary parameters, and bold face $\pvecz$ shows all the hyperparameter of population model including redshift.
\subsection{Individual Event Inference}
\label{sec:indiv_event_inference}

We employ Hierarchical Bayesian Modeling (HBM) to constrain a population model with
gravitational wave data \cite{ThraneTalbot2019}. A total of $N$ discrete detections with a mixture of merger types i.e. BBH merger, NSBH mergers or BNS mergers. Those detections provide
merger data denoted as $d_1,d_2,d_3,\cdots,d_N$.
Each stretch of data $d_j$ for \(j=1,2,\cdots,N\) is used to infer the properties of the
associated event with that data segment. We also refer to it as the likelihood function
$\ell_j(\svecz)\equiv p(d_j|\svecz)$
of a source, often evaluated using matched filtering against a template bank of waveform models. When calculated in full with strain data and a waveform model, the full likelihood function expresses the probability of a specific waveform model with parameters
\(\svecz\) in the data $d_j$.  We may use a uniform or informative reference prior $\pi(\svecz)$ for finding a posterior probability using the Bayes theorem as
given in Equation \eqref{eq:Bayes_ind},

\begin{equation}
	\label{eq:Bayes_ind}
	p(\svecz|d_j) \propto p(d_j|\svecz) \cdot \pi(\svecz).
\end{equation}

This posterior probability constrain the properties of each
binary, such as mass, spin, eccentricity, distance, and sky location. \revision{In practice after performing the parameter estimation of each event against the desired waveform model, we get the discrete samples which represent the likelihood $\ell_j(\svecz)$ of that event. Once we have the likelihood of all the desired events, we use them to inform our population analysis explained in the following section.}

\subsection{Population Inference}
\label{sec:pop_model_inference}

Given the likelihood $\ell(\svecz)$ of individual events and their reference prior $\pi(\svecz)$, we proceed with a hierarchical Bayesian framework given in Equation \eqref{eq:Bayes_pop} to infer the population-level parameters $\pvecz_i$ given the dataset $\data=\{{d_j}\}_{j=1}^N$.

\begin{align}
	\label{eq:Bayes_pop}
	\!p\left(\pvecz_i | \data \right)
	 & = \frac{
		\pi(\pvecz_i)\,
		p(\data | \pvecz_i)
	}{
		p\left( \data \right)
	},
\end{align}
where $p(\pvecz_i| \data)$ is posterior distribution of hyperparameter $\pvecz_i$, $\pi(\pvecz_i)$ is the population prior, and $\mathcal{L}(\pvecz_i)\equiv p(\data | \pvecz_i)$ is the population likelihood, a core integral for population inference. The term $p(\data)$, known as Bayesian evidence, serve as normalization constant and often omitted in sampling-based inference.
Therefore, in practice we will use the likelihood function $\mathcal{L}(\pvecz_i)$ to compute the posterior distribution $p(\pvecz_i| \data) \propto \mathcal{L}(\pvecz_i)\cdot \pi(\pvecz_i)$ of the population parameters $\pvecz_i$.

To conduct our analysis we have used the Inhomogeneous Poisson Process \cite{2019MNRAS.486.1086M, 2004AIPC..735..195L, 2015PhRvD..91b3005F} for each type of population
\begin{equation}
	\label{eq:likelihood}
	\mathcal{L}(\pvecz) \propto
	e^{-\mu{(\pvecz)}}
	\prod_{j=1}^N
	\int\ell_j(\svecz) \cdot \comp(\svecz\mid\pvecz)
	\sqrt{ g_{\svecz}}
	d \svecz,
\end{equation}
where $g_{\svecz}$ is the determinant of the
metric over those coordinates, and  $\comp$ is the merger rate density in source frame of reference (Equation \eqref{eq:total_rate}), and $\ell_j(\svecz)$ is the likelihood of individual event. For source parameters, we adopt a usual uniform metric over all intrinsic and extrinsic
parameters, such that $\sqrt{g_{\svecz}}d\svecz = T_{\mathrm{obs}} \times dz (1+z)^{-1} (dV_c/dz) \times dm_1 dm_2 \times $  appropriate factors for
spin which depend on the coordinate representation adpoted for them.   The term $\ell_j(\svecz)=p(d_j|\svecz)$ is
the likelihood of individual events and can be read from the data files (real or synthetic data). The exponent $\mu{(\pvecz)}$ in population likelihood is the total expected number of detections under
the given population parametrization $\pvecz$, the complete expression is given in Equation \eqref{eq:mu}.

These calculations are analytically intractable and must be performed numerically. Specifically we have used Normalizing Flow enhanced Metropolis adjusted Langevin algorithm \cite{2022PNAS..11909420G}, provided by \texttt{flowMC} \cite{git_flowMC,2023JOSS....8.5021W,2022PNAS..11909420G}. Further technical details are given in appendix \ref{subsec:appendix:hbm}.

\subsubsection{Expected Rate Estimation}
\label{subsection:volume}

The expected number of GW detections can be formulated as an integral over the intrinsic source parameter space $\svec$ and redshift $z$ modulated by an appropriate selection (weighting) function. The total expected number of detections summing over all populations is given by

\begin{align}
	\label{eq:mu}
	\mu(\pvecz)= \int P_{\mathrm{det}} (\svec;z)\cdot \comp(\svecz\mid\pvecz)\sqrt{ g_{\svecz}}
	d \svecz.
\end{align}

Here $P_{\mathrm{det}}(\svec;z)$ is the detection probability for a source with intrinsic parameters $\svec$ at redshift $z$. The probability of detection $P_{\mathrm{det}}(\svec;z)$ is the fundamental ingredient in the calculation of the
expected number of detections $\mu(\pvecz)$. This can be
provided in multiple ways, such as injection-based
\cite{2022arXiv220400461E, ligo_scientific_collaboration_and_virgo_2023_7890398},
analytical \cite{2021ApJ...922..258V}, semi-analytical model with
fixed threshold \cite{1993PhRvD..47.2198F,2023PhRvD.108d3011E, 2020LRR....23....3A}
and recent development of training neural nets on real injections
\cite{2024arXiv240816828C}. We can choose any of the methods based on
analysis requirements and computational resources. However, for
illustration purposes, we have explained the semi-analytical approach and its
approximation with neural net in the following subsections.

\subsubsection{Semi Analytical Approach for Detection Probability}
\label{subsubsection:semi_analytical_volume}

We estimate the detection probability $P_{\mathrm{det}}(\svec;z)$ of a source at redshift $z$ using a semi-analytical approach that combines numerical evaluation of signal-to-noise(SNR) over a population sources with theoretical detector sensitivity, as characterized through one-sided power spectral density (PSD) $S_n(f)$ curves given in \texttt{LALSimulation} package \cite{lalsuite, 2020SoftX..1200634W} such as for Advanced LIGO \cite{ligo_scientific_collaboration_and_virgo_2023_7890398} or Virgo \cite{2015CQGra..32b4001A}. Similarly, waveforms $h(f|\svec)$ for a source with intrinsic parameter $\svec$ at redshift $z$ are modeled using frequency-domain approximants from the \texttt{LALSimulation} package.

To compute the matched-filtered SNR $\rho_{\mathrm{opt}}$ for a GW signal $h(f|\svec)$, we generate synthetic injections of binary systems with source-frame parameters $\svec$ sampled uniformly over the desired range and redshift $z$ being fixed or evolving for all sources to compute their waveforms.
After having the waveform and simulated noise curve, we compute the SNR using the following equation

\begin{align}\label{eq:snr}
	\rho_{\mathrm{opt}}^2 = 4 \int_{f_{\mathrm{min}}}^{f_{\mathrm{max}}} \frac{|h(f|\svec)|^2}{S_n(f)}df.
\end{align}

For default calculations, the SNR is computed with a lower frequency cutoff $f_{\mathrm{min}} = 10\,\mathrm{Hz}$, an upper cutoff $f_{\mathrm{max}} = 2048\,\mathrm{Hz}$, a reference frequency of $f_{\mathrm{ref}} = 40\,\mathrm{Hz}$. The PSD used is \texttt{SimNoisePSDaLIGO175MpcT1800545}, and the waveform model is \texttt{IMRPhenomPv2}. We use the \texttt{SimInspiralChooseFDWaveform} interface for waveform generation, and the \texttt{Planck15} cosmology for computing luminosity distance $d_L$. For a compact binary at luminosity distance $d_L$, $h(f)\propto1/d_L$, and the SNR is inversely proportional to the distance, $\rho(z)= \rho_0(z=0)/d_L(z)$.

After computing the optimal SNR ($\rho_{\mathrm{opt}}$) of injected sources. The detection probability $P_{\mathrm{det}}(\svec;z)$ is then estimated using an empirical fit calibrated against orientation-averaged Monte Carlo simulations. Specifically, we use the dimensionless ratio $w = \rho_{\mathrm{thresh}} / \rho_{\mathrm{opt}}$, where $\rho_{\mathrm{thresh}}=8$ is the detection threshold chosen for this study. A source is considered detectable if $w < 1$. The detection probability $P_{\mathrm{det}}(\svec;z)$ is evaluated using the following analytical approximation given in appendix A \cite{2015ApJ...806..263D}:
\begin{align}
	\label{eqn:p_det}
	P_{\mathrm{det}}(\svec;z) = & a_2 (1 - w)^2 + a_4 (1 - w)^4 + a_8 (1 - w)^8 \nonumber \\
	                            & + (1 - a_2 - a_4 - a_8)(1 - w)^{10},
\end{align}
where the coefficients are $a_2 = 0.374222$, $a_4 = 2.04216$, and $a_8 = -2.63948$.

% Finally, the characteristic volume, where the source can be detected is calculated by integrating the detection probability $P_{\mathrm{det}}$ over the redshift of the source, given in equation \eqref{eq:volume}.

% \begin{equation}
%     \label{eq:volume_gen}
%     V(\svec) = \int P_{\mathrm{det}}(\svec,z) \cdot p(z|\zvec=0) dz,
% \end{equation}
% one can expand the redshift distribution $p(z|\zvec=0)$ as

% \begin{equation}
%     \label{eq:volume}
%     V(\svec) = \int \frac{dV_c}{dz}\frac{P_{\mathrm{det}}(\svec)}{1+z}  dz .
% \end{equation}

% Here $P_{\mathrm{det}}$ is the detection probability for a source with intrinsic parameters $\svec$, at redshift $z$. The term $dV_c/dz$ is the differential comoving volume, and $(1+z)^{-1}$ accounts for the time dilation effect.

\subsubsection{Neural Net Estimator for Detection Probability}
\label{subsubsection:neural_net_pdet}
\begin{figure*}
	\includegraphics[width=\textwidth]{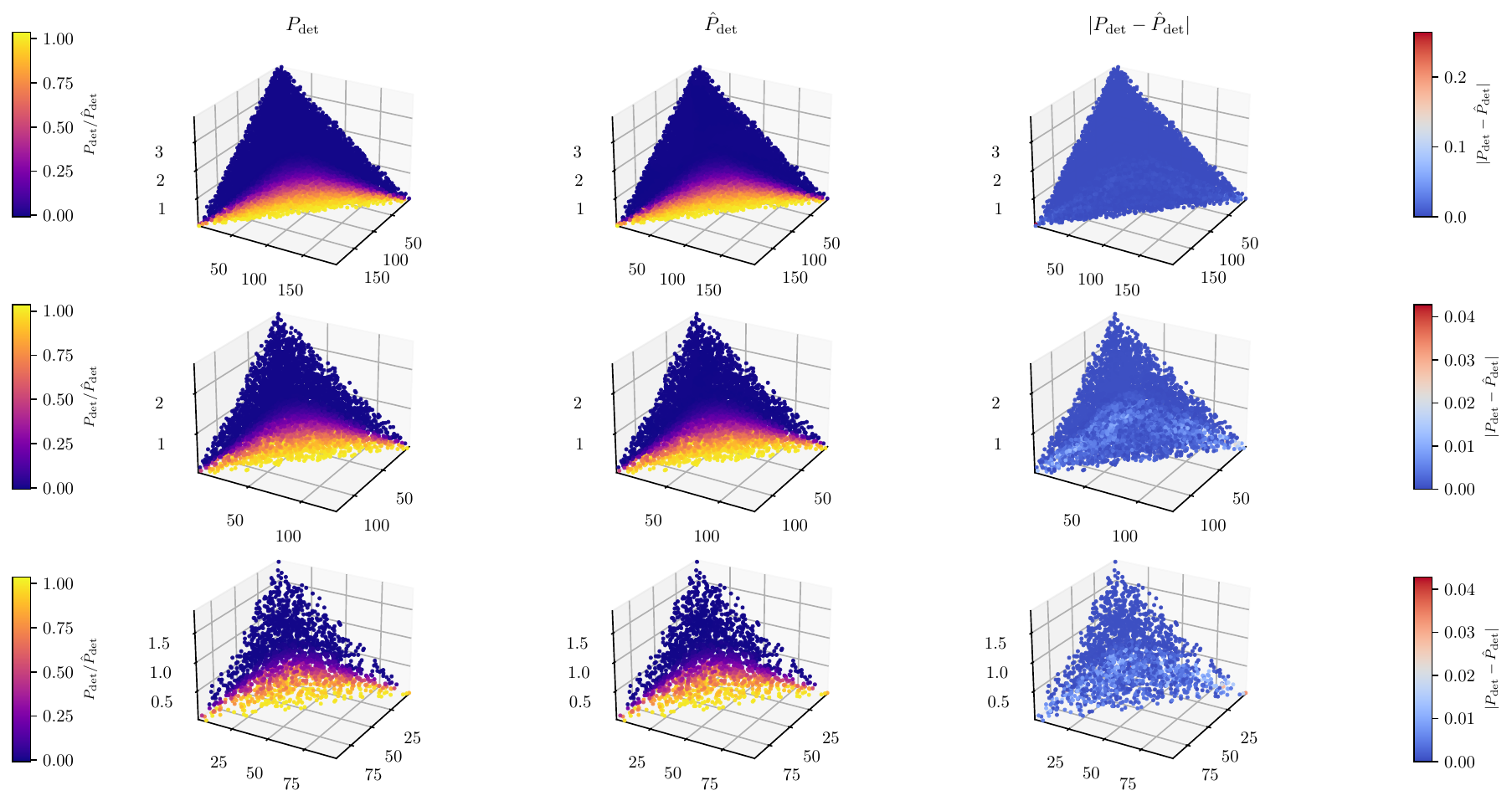}
	\caption{\textbf{Error Between Semi-Analytical and Neural Net Estimators:} In each plot primary mass, secondary mass and redshift are plotted on x,y and z axis respectively. For each row have taken a cut the uniformly scattered points by a hyperplane to show different values of $P_{\mathrm{det}}$. The L.H.S. plot shows the semi-analytical $P_{\mathrm{det}}$ calculations on randomly generated injections from uniform distribution. The middle figure shows $\hat{P}_{\mathrm{det}}$ neural net estimate of $P_{\mathrm{det}}$, and R.H.S. shows the error between them.}
	\label{fig:pdet}
\end{figure*}

Interpolating $P_{\mathrm{det}}$ values during inference is computationally expensive. Therefore,  to overcome this, we use a Deep Multi-Layer Perceptron (DMLP) \cite{rosenblatt1958perceptron} to estimate $P_{\mathrm{det}}$ of the sources during inference. The input layer has one neuron \cite{mcculloch1943logical} per $P_{\mathrm{det}}$ parameter (intrinsic and extrinsic), and the output layer has one neuron for the $P_{\mathrm{det}}$. The model is trained using backpropagation \cite{rumelhart1986learning,linnainmaa1976taylor} with ReLU activation \cite{agarap2018deep} between layers. Figure~\ref{fig:pdet} shows the semi-analytical $P_{\mathrm{det}}$, which we compute using the method described in previous subSection \ref{subsubsection:semi_analytical_volume}, the trained neural $P_{\mathrm{det}}$, which we actually use for inference and error between them. For this study, we prefer to train $P_{\mathrm{det}}$ on a uniform distributed sources as compare to training the volume, because, we see a better performance of neural net training on $P_{\mathrm{det}}$ as compared to volume values.

\section{Validation Studies}
\label{sec:validation_analyses}

To validate the \gwk, we reproduced the two previously published studies: non-spinning eccentric population \cite{2024PhRvD.110f3009Z} and BBH mass distribution using \revision{fourth Gravitational-Wave Transient Catalog (GWTC-4) population \cite{2025arXiv250818083T}}. We have also made the injection recovery by generating posteriors with \gwk. We have generated two synthetic populations: first, is the spinning eccentric BBH and second is the circular mixture population of BBHs, BNS, NSBH based on the multi-source model, detailed in Appendix C3 of \cite{2019PhRvX...9c1040A}. The conventional methods used to generate synthetic data for this work are summarized in the Appendix \ref{subsection:syn-pop-uncertainties}.

\subsection{Reproduced Published Results: Non-Spinning Eccentric BBHs}
\label{subsec:reproduced_eccentric}

We have reproduced the optimistic case of the eccentricity distribution $(\sigma_\epsilon=0.15)$ as presented in
\cite{2024PhRvD.110f3009Z}. We have taken the same dataset and VT which was used in the original study and made the
analysis using \gwk. The only thing we changed in this analysis for \gwk is that we trained the VT as described in Section \ref{subsection:volume} to accelerate the analysis which also shows that using the neural net approach for VT still gives us consistent results with the previous study.

\begin{figure}[ht!]
	\includegraphics[width=\columnwidth]{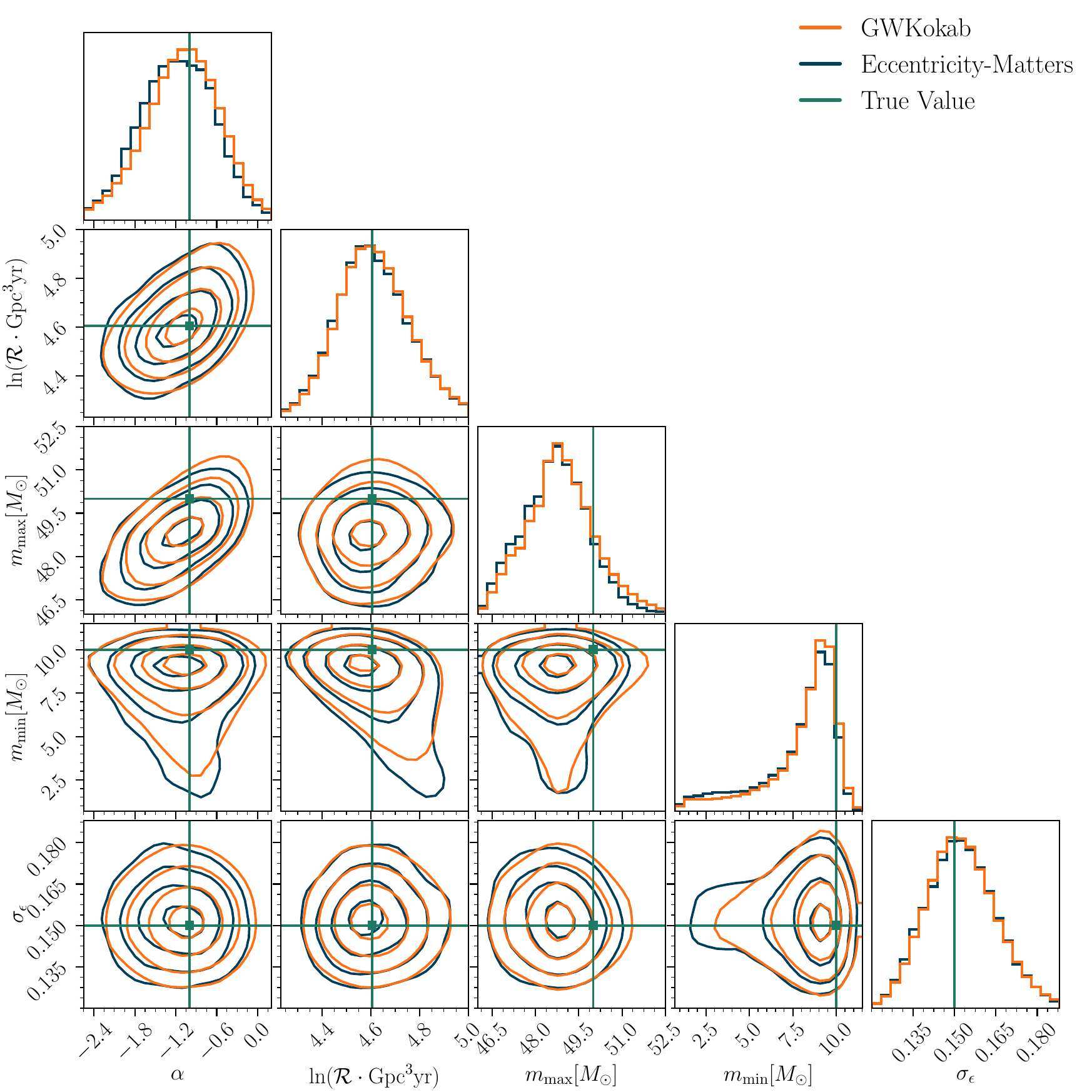}
    \caption{\revision{\textbf{Non-Spinning Eccentric BBHs:} Corner plot showing population parameter recovery for the dataset used in \cite{2024PhRvD.110f3009Z}. \gwk achieves consistent inference at 2\% of the original study's computational cost.}\label{fig:ecc_corner}
	}
\end{figure}

\revision{Figure~\ref{fig:ecc_corner} shows the recovery of population parameters with \gwk and previously published work, it demonstrates good agreement. The machine learning based sampler \textsc{flowMC}, neural net based VT approximation and GPU based inference give us significant boost in computational efficiency while keeping the consistent science results.}
This analysis was completed in 0.14 hours (equivalent to 8.44 minutes) using \gwk, in contrast to 10.41 hours (equivalent to 625.12 minutes) on the \textsc{Ecc-Matters} \cite{git_eccmatters} framework. These results demonstrate that \gwk is capable of accurately recovering population parameters while significantly reducing computational costs. Specifically, the computational time was reduced by approximately 98\% for the same analysis on the same machine.

\subsection{Injection Recovery: Spinning Eccentric BBHs}
\label{subsec:spinning_eccentric}

We conducted a straw-man analysis to demonstrate the ability of \gwk to recover the population parameters of spinning eccentric BBHs. We have used power law in mass $m_1,q$ detailed in Section \ref{appendix:subsec:population-models}, truncated gaussian for aligned spin $\chi_{i,z}$, and a half-normal distribution for eccentricity $\epsilon$ parametrized by a single width parameter as used in the previous Section \ref{subsec:reproduced_eccentric}.

\begin{figure}[t!]
	\includegraphics[width=\columnwidth]{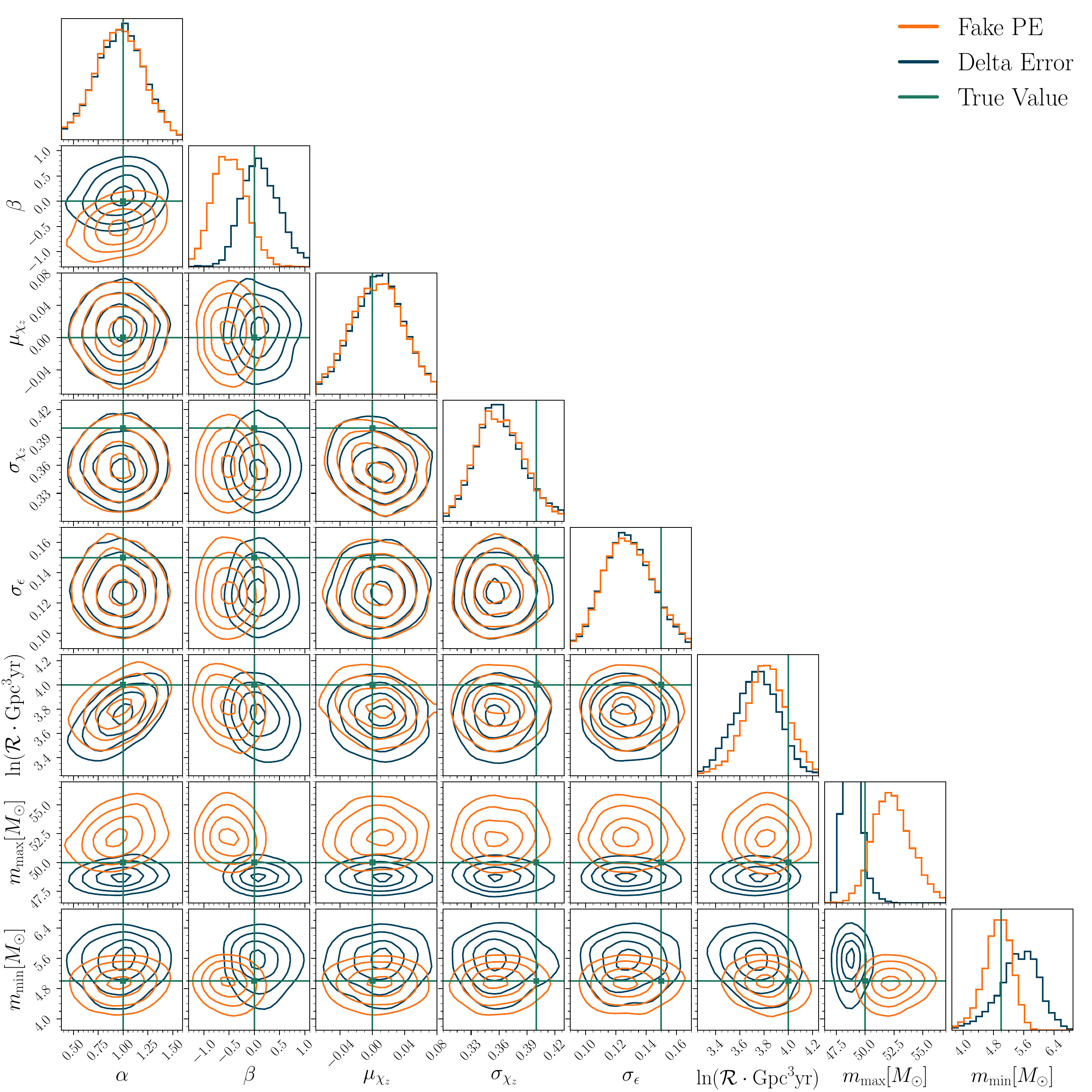}
	\caption{\revision{\textbf{Spinning Eccentric Population:} Corner plot for population model hyperparameters $\pvec$. Results using synthetic PEs (Appendix \ref{subsection:syn-pop-uncertainties}) are shown in {\color[HTML]{fd7014}dark orange}, while delta error PE recoveries are in {\color[HTML]{003f5c}dark slate gray}. {\color[HTML]{1f7a63}Teal lines} indicate true parameter values.
	% The corner plot of the recovered population model hyperparameters $\pvec$. The two different colors show the different error types in the true values. The {\color[HTML]{fd7014}dark orange} one is showing the recovery using fake PEs described in appendix \ref{subsection:syn-pop-uncertainties}, {\color[HTML]{003f5c}dark slate gray} one is showing the recovery using delta error PEs, and lines in {\color[HTML]{1f7a63}teal} shows their true values.
	}\label{fig:SEBBH_corner}}
\end{figure}

Table \ref{tab:SEBBH} provides the priors adopted on our analysis of these synthetic observations, as well as our specific choices for synthetic model parameters.
We have generated the injections using the same model and applied the realistic VT effects on mass and spin keeping the eccentricity independent by using previously \cite{2019PhRvD.100d3012W, pop-models-aps-2021-vt} generated semi-analytical VT based on the PSD \textsc{PSDaLIGO140MpcT1800545} \cite{Barsotti2018} and the calibration with real O3 sensitivity injections using the least square method. Additionally, to make it computationally efficient we trained this calibrated VT using the methodology described in Section \ref{subsection:volume}.
As in the previous study, we generate synthetic data using the method described in the Appendix
\ref{subsection:syn-pop-uncertainties}, based on the approach described in Section III.A of \cite{2024PhRvD.110f3009Z}.
Extending the synthetic data approach adopted in the previous analysis, the spin and eccentricity parameters are assumed to have Gaussian measurement errors with a characteristic one-dimensional standard deviation of $0.1$. The injections and their corresponding fake posteriors used for this analysis are shown in Figure~\ref{fig:fake_pe} for mass and spin. To further validate the method, delta function like errors are also employed, demonstrating that the code performs correctly in the idealized case. In this delta-error setup, we assign a uniform distribution centered on the true value of each parameter with a very narrow width, effectively mimicking a delta function: for the masses we used $1 \Ms$, while for the spins and eccentricity we used $0.1$. For both the delta-error and fake-PE cases, the same set of injections was used, with 5000 samples per event.
As in previous work, these parameter uncertainties are adopted for simplicity in validating our algorithm; they are not intended to reproduce fully realistic posteriors, particularly as correlations and mass dependence are omitted. Figure~\ref{fig:SEBBH_corner} presents the results of our inference, expressed as posterior distributions of the model hyperparameters $\pvec$. For comparison, the true properties of the underlying synthetic population are also shown. As expected, our model successfully recovers the synthetic population properties. These results demonstrate that \gwk can reliably recover the population parameters of spinning, eccentric BBHs.

% explanation: following lines are mentioned in figure caption
% The two different colors in Figure \ref{fig:SEBBH_corner} indicate the error types in the recovery: the blue curves correspond to the fake PEs described in Appendix \ref{subsection:syn-pop-uncertainties}, while the orange curves correspond to the delta-error PEs.

\begin{table}[t!]
	\centering
	\begin{tabular}{c|cccccccc}
		\hline
		Parameter
		 & $\ln\srate_0$
		 & $\alpha$
		 & $\beta$
		 & $m_{\mathrm{min}}$
		 & $m_{\mathrm{max}}$
		 & $\mu_{\chi_{z}}$
		 & $\sigma_{\chi_{z}}$
		 & $\sigma_\epsilon$   \\
		\hline
		Synthetic Value
		 & 4
		 & 1
		 & 0
		 & 5
		 & 50
		 & 0
		 & 0.4
		 & 0.15                \\
		\hline
		Low
		 & 0
		 & $-6$
		 & $-6$
		 & 1
		 & 30
		 & $-1$
		 & 0
		 & 0                   \\
		\hline
		High
		 & 10
		 & 6
		 & 6
		 & 20
		 & 80
		 & 1
		 & 1
		 & 1                   \\
		\hline
	\end{tabular}
	\caption{\textbf{Spinning Eccentric Population:} True parameters to generate synthetic population and priors used for Bayesian inference. The ``Low" and ``High" indicate the lower and upper bounds of the uniform prior, respectively.}
	\label{tab:SEBBH}
\end{table}

\subsection{Injection Recovery: Multi-Source Population with Independent Rate Parameters}
\label{subsec:multi_source}

\revision{To demonstrate our ability to reconstruct a mixture of multiple plausible subpopulations, we generate and recover
	subpopulations of BBH, BNS, and NSBH with their independent merger rates}. Specifically, we used multi-source model, as outlined and explained in Section III-C-3 and Appendix B-4-a of \cite{2019PhRvX...9c1040A} respectively. In this model, the BBH population consists of one power law model superimposed with one gaussian distribution; the BNS is characterized by a truncated gaussian; and the NSBH is similarly characterized by another
truncated gaussian.
\revision{We recover the boundary parameters $m_{\mathrm{min,pl,BBH}}$ and $m_{\mathrm{max,NSBH,BH}}$ from the data, while $m_{\mathrm{min,peak,BBH}}$ is set to follow $m_{\mathrm{min,pl,BBH}}$. The following parameters are fixed: $m_{\mathrm{max,pl,BBH}} = m_{\mathrm{max,peak,BBH}} = 100$, $m_{\mathrm{min,NSBH,BH}} = 5$, $m_{\mathrm{min,BNS}} = m_{\mathrm{min,NSBH,NS}} = 1$, and $m_{\mathrm{max,BNS}} = m_{\mathrm{max,NSBH,NS}} = 3$.
	All the BNS, including NSs in NSBHs have the same mass distribution with parameters $\mu_{m,\mathrm{BNS}}$ and $\sigma_{m,\mathrm{BNS}}$.}

For the subpopulations spin, BHs (powerlaw, peak, and NSBH) have an independent default spin model for their primary and secondary components, with $\zeta\equiv1$, as described in Appendix B-2-a \cite{2019PhRvX...9c1040A}. Similarly, all the NSs (within BNS and NSBH) have the same independent default spin model, with $\zeta\equiv0$. To reduce the computational cost, we have given a same spins to primary and secondary components of binaries, however we can can give different spins to both components. We also prefer truncated gaussian for the spin distribution as compare to beta distribution used in the original study because gaussian is well behaved and easy to recover with auto-differentiation. The eccentricity is fixed to zero for all the subpopulations and redshift evolution is also ignored for simplicity. However, \gwk have the flexibility to add eccentricity and redshift evolution in the model.

\begin{figure*}[ht!]
	\includegraphics[width=0.48\textwidth]{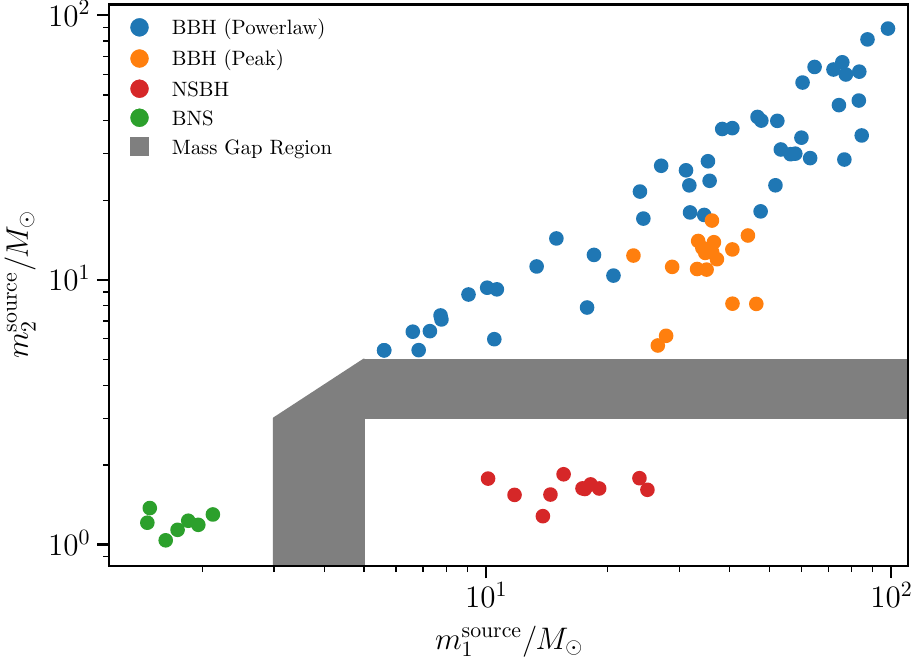}
	\includegraphics[width=0.48\textwidth]{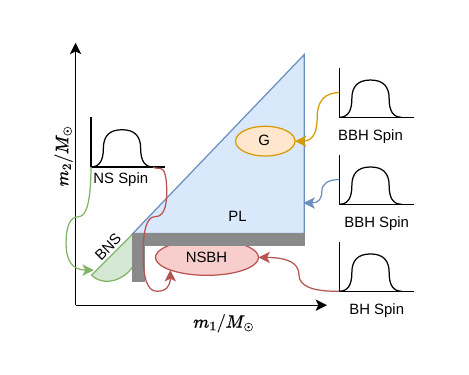}
	\caption{\revision{\textbf{Multi-source population overview:} The left panel illustrates the population injections for BBH power law ({\color[HTML]{1f77b4}blue}), BBH peak ({\color[HTML]{ff7f0e}orange}), BNS ({\color[HTML]{2ca02c}green}), and NSBH ({\color[HTML]{d62728}red}) with their independent rates. These are represented schematically in the right panel, which outlines the separate spin and rate parameters for each category. Notably, the spin distribution for neutron stars is consistent across both BNS and NSBH models. The {\color[HTML]{7f7f7f}grey} region indicates the mass gap ($3\text{--}5 \, \Ms$).}\label{fig:multisource_inj}}
\end{figure*}

\begin{figure}[ht!]
	\includegraphics[width=\columnwidth]{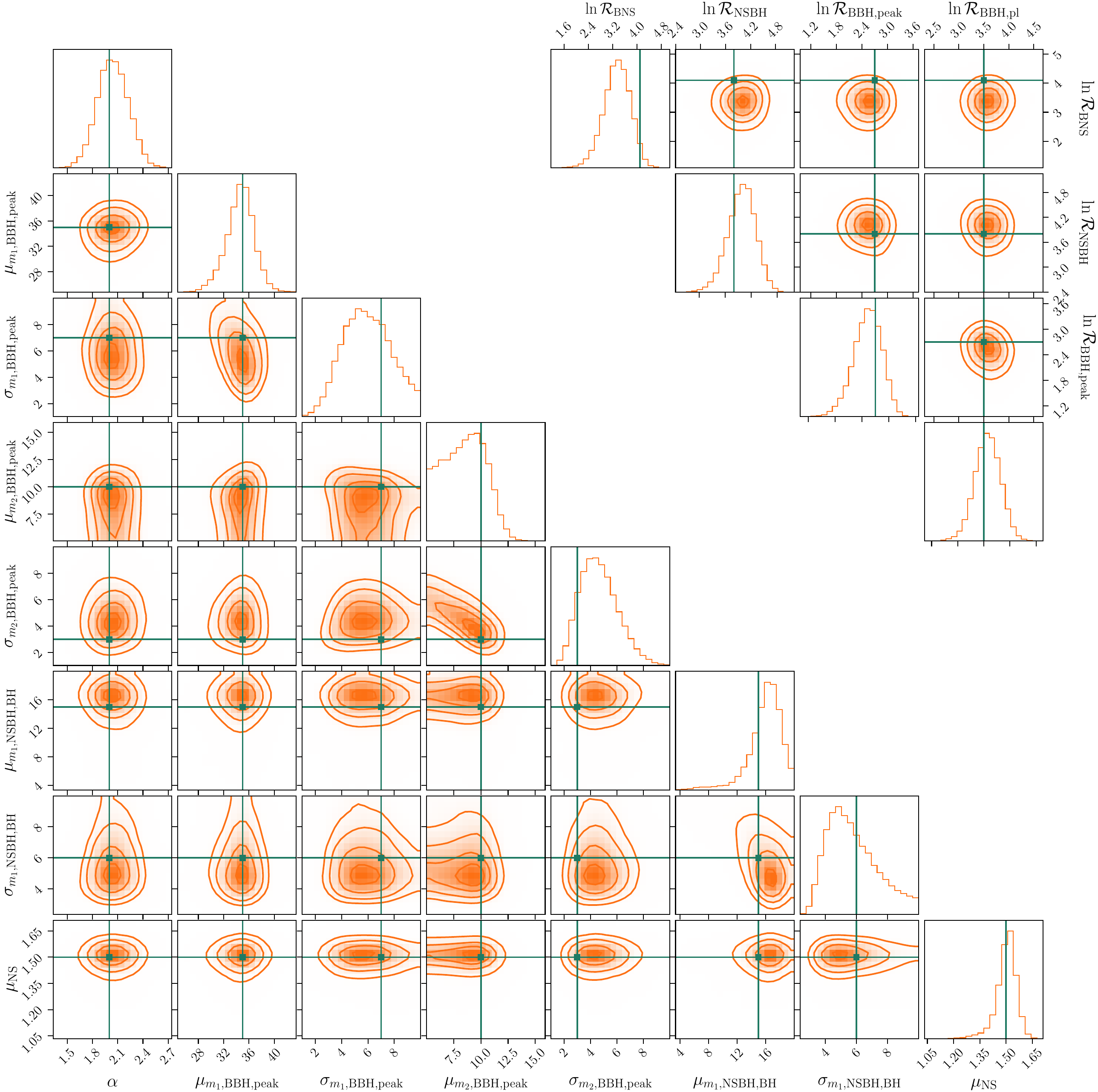}
	\caption{\revision{
			\textbf{Posterior corner plots for the multi-source population model:} The lower panels show the recovery of the model's shape hyperparameters, while the upper panels display the independent rate parameters for the BNS, NSBH, and BBH subpopulations. Contours represent the 68\% and 95\% credible intervals.
		}\label{fig:sub_pop_corner}}
\end{figure}

\begin{figure}[ht!]
	\includegraphics[width=\columnwidth]{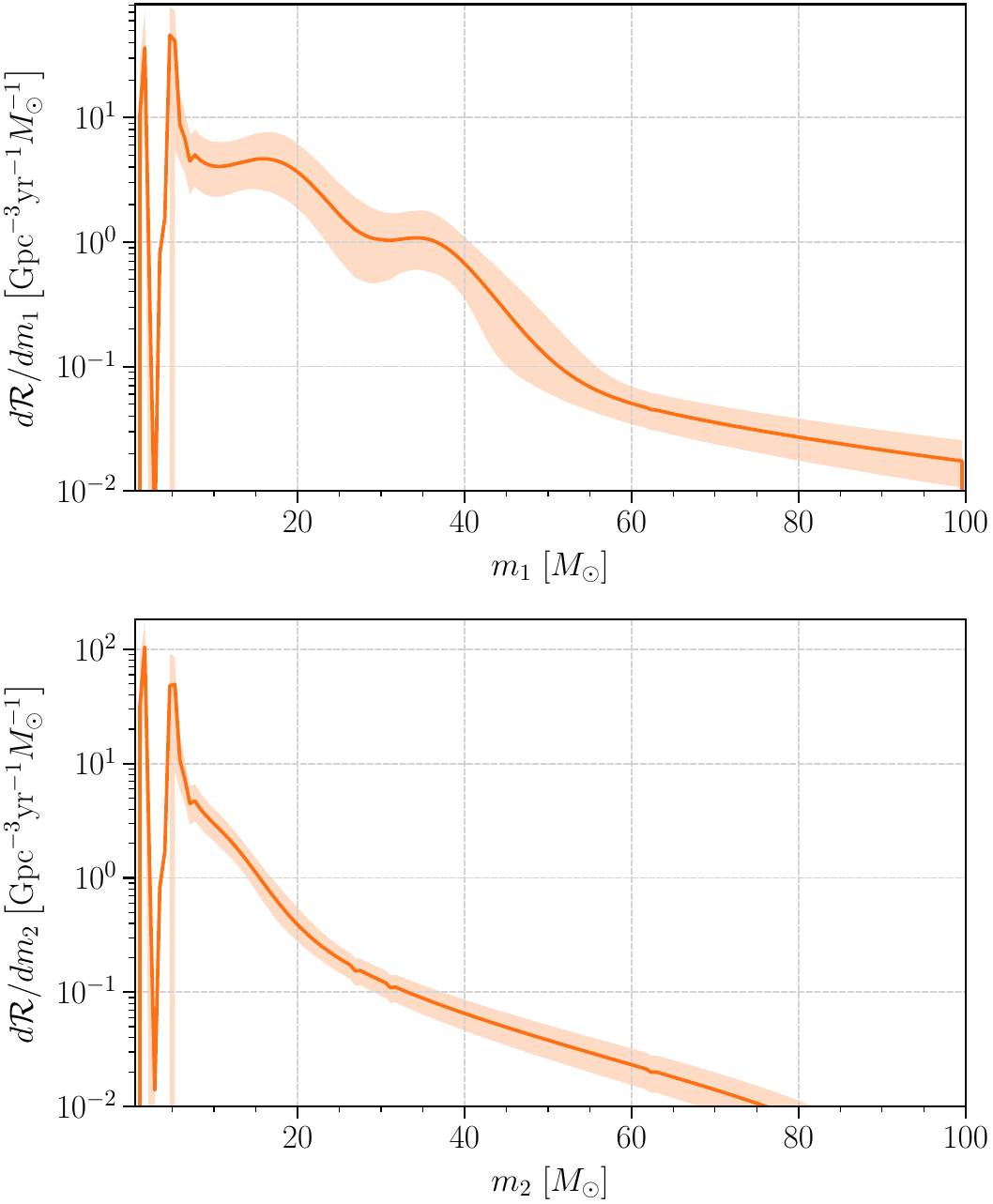}
	\caption{
		\revision{\textbf{Multi-Source Population:} Upper and lower plots show PPDs (Appendix \ref{subsec:ppd}) for primary and secondary mass, respectively.}\label{fig:mass_ppd}
	}
\end{figure}

% In our analysis, we adopt precisely the same model family and precisely the same priors as described in table XIV of \cite{2019PhRvX...9c1040A}, as summarized in Table \ref{tab:MSP}.
We generate a synthetic population of compact
binaries, illustrated in Figure~\ref{fig:multisource_inj} using the specific model parameters provided in Table \ref{tab:MSP}. Synthetic detections are identified using
the same VT model used in previous Section \ref{subsec:spinning_eccentric}.
As previously, we employ the naive procedure described in the Appendix to generate synthetic observational errors.
Figure~\ref{fig:sub_pop_corner} demonstrates that \gwk recovers properties of the underlying model; for brevity, we omit a full hyperparameter corner plot.
Figure~\ref{fig:mass_ppd} demonstrates that our model recovers the one-dimensional rate distributions.

These results demonstrate that \gwk is highly efficient in recovering the properties of subpopulations with independent rates. We can see in Figure~\ref{fig:mass_ppd} that PPDs are reflecting the properties of the injected population. This multi-source approach and separate PPDs of Primary and Secondary Mass distributions has the potential to reveal important insights into the underlying population characteristics such as the mass gap and the presence of subpopulations, which can be crucial for understanding the formation and evolution of compact binary systems.

\begin{table}[t!]
	\centering
	\begin{tabular}{c|c|c}
		\hline
		Parameter                                                                & Synthetic Value & Priors                                \\ \hline
		$\ln{\mathcal{R}_{\mathrm{BBH,pl}}}$                                     & 3.5             & \multirow{4}{*}{$\mathcal{U}(0, 10)$} \\
		$\ln{\mathcal{R}_{\mathrm{BBH,peak}}}$                                   & 2.7             &                                       \\
		$\ln{\mathcal{R}_{\mathrm{NSBH}}}$                                       & 3.8             &                                       \\
		$\ln{\mathcal{R}_{\mathrm{BNS}}}$                                        & 4.1             &                                       \\ \hline
		$\alpha$                                                                 & 2               & $\mathcal{U}(-4, 12)$                 \\ \hline
		$m_{\mathrm{min,pl,BBH}}$                                                & 5               & $\mathcal{U}(3, 10)$                  \\ \hline
		$\mu_{m_1,\mathrm{peak,BBH}}$                                            & 35              & $\mathcal{U}(20, 50)$                 \\ \hline
		$\mu_{m_2,\mathrm{peak,BBH}}$                                            & 10              & $\mathcal{U}(5, 40)$                  \\ \hline
		$\sigma_{m_1,\mathrm{peak,BBH}}$                                         & 7               & \multirow{3}{*}{$\mathcal{U}(1, 10)$} \\
		$\sigma_{m_2,\mathrm{peak,BBH}}$                                         & 3               &                                       \\
		$\sigma_{m,\mathrm{NSBH,BH}}$                                            & 6               &                                       \\ \hline
		$m_{\mathrm{max,NSBH,BH}}$                                               & 35              & $\mathcal{U}(30, 70)$                 \\ \hline
		$\mu_{m,\mathrm{NSBH,BH}}$                                               & 15              & $\mathcal{U}(3, 20)$                  \\ \hline
		$\mu_{m,\mathrm{BNS}}$                                                   & 1.5             & $\mathcal{U}(1, 3)$                   \\ \hline
		$\sigma_{m,\mathrm{BNS}}$                                                & 0.25            & $\mathcal{U}(0.05, 1)$                \\ \hline
		$\sigma_{\chi_{z},\mathrm{BBH,pl}}$                                      & 0.3             & \multirow{3}{*}{$\mathcal{U}(0, 1)$}  \\
		$\sigma_{\chi_{z},\mathrm{BBH,peak}}$                                    & 0.4             &                                       \\
		$\sigma_{\chi_{z,1},\mathrm{NSBH,BH}}$                                   & 0.6             &                                       \\ \hline
		$\sigma_{t,\mathrm{BBH,pl}}$                                             & 1.5             & \multirow{3}{*}{$\mathcal{U}(0, 4)$}  \\
		$\sigma_{t,\mathrm{BBH,peak}}$                                           & 2               &                                       \\
		$\sigma_{t,1,\mathrm{NSBH,BH}}$                                          & 2.5             &                                       \\ \hline
		$m_{\mathrm{min,peak,BBH}}$                                              & 5               & $m_{\mathrm{min,pl,BBH}}$             \\
		$m_{\mathrm{max,peak,BBH}}$                                              & 100             & $m_{\mathrm{max,pl,BBH}}$             \\
		$\mu_{m,\mathrm{NSBH,NS}}$                                               & 1.5             & $\mu_{m,\mathrm{BNS}}$                \\
		$\sigma_{m,\mathrm{NSBH,NS}}$                                            & 0.25            & $\sigma_{m,\mathrm{BNS}}$             \\ \hline
		$\beta$                                                                  & 0.7             & \multirow{7}{*}{Fixed}                \\
		$m_{\mathrm{max,pl,BBH}}$, $m_{\mathrm{max,peak,BBH}}$                   & 100             &                                       \\
		$m_{\mathrm{min,NSBH,BH}}$                                               & 5               &                                       \\
		$m_{\mathrm{min,NSBH,NS}}$, $m_{\mathrm{min,BNS}}$                       & 1               &                                       \\
		$m_{\mathrm{max,NSBH,NS}}$, $m_{\mathrm{max,BNS}}$                       & 3               &                                       \\
		$\sigma_{\chi_{z},\mathrm{BNS}}$, $\sigma_{\chi_{z,2},\mathrm{NSBH,NS}}$ & 0.1             &                                       \\
		$\sigma_{t,\mathrm{BNS}}$, $\sigma_{t,2,\mathrm{NSBH,NS}}$               & 1               &                                       \\ \hline
	\end{tabular}
	\caption{\revision{\textbf{Multi-Source Population Parameters and Priors:} Summary of the true hyperparameter values used to generate the synthetic population and the corresponding prior distributions. $\mathcal{U}(a,b)$ represents a uniform distribution. Prior entries containing other parameter names indicate duplicated parameters that are constrained to have the same value during inference. Parameters marked as Fixed are kept constant at their synthetic values. The \textrm{pl} shows the powerlaw component of the multi-source model and \textrm{peak} shows the gaussian component of the multi-source model, full description of the model parameters can also been see in \cite{2019PhRvX...9c1040A}.}}
	\label{tab:MSP}
\end{table}

\subsection{Reproduced Published Results: GWTC-4 BBHs Population}

\begin{figure}[t!]
	\includegraphics[width=\columnwidth]{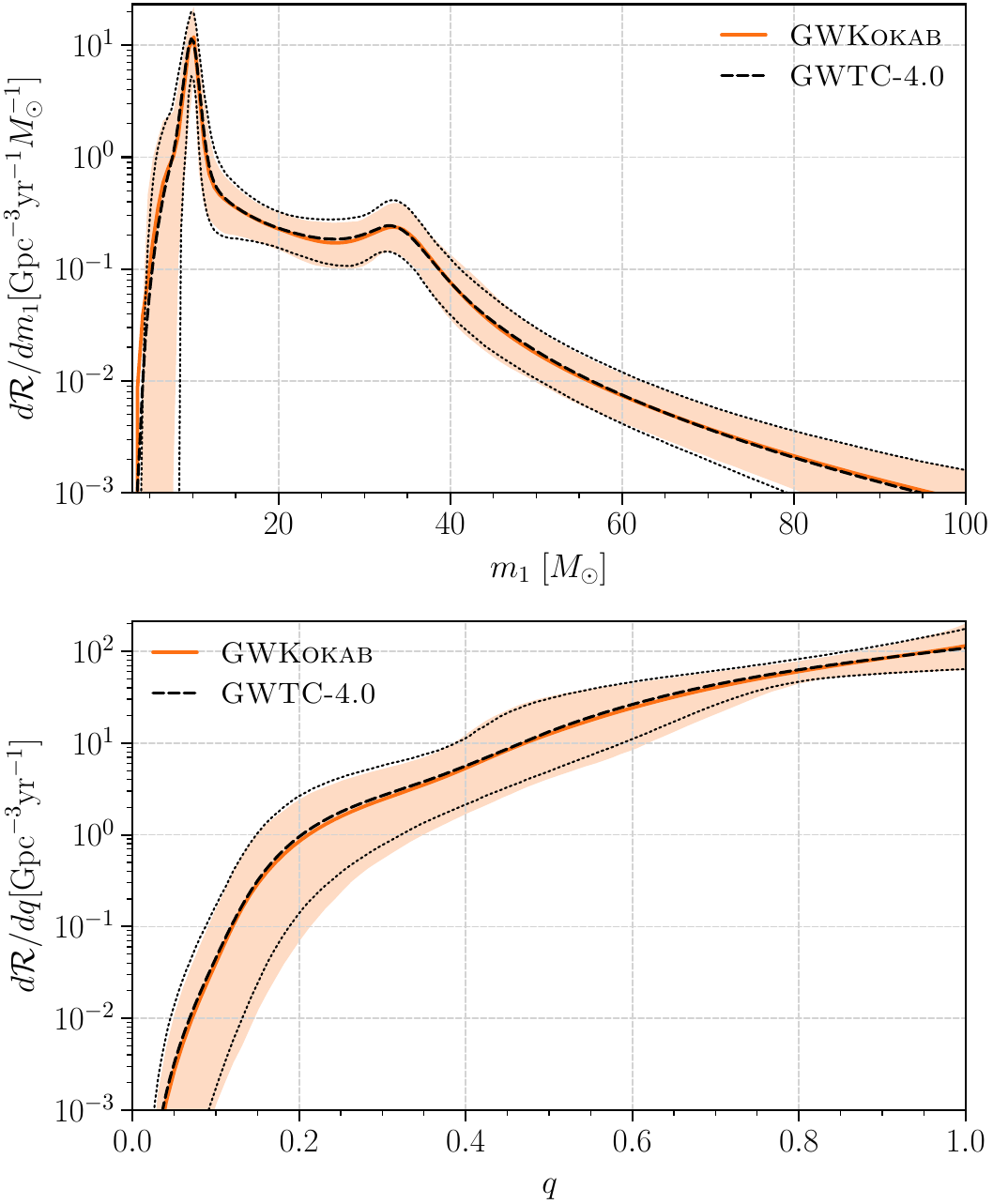}
	\caption{\textbf{GWTC-4 Reproduction (Masses):} Primary mass (top) and mass-ratio (bottom) distributions for GWTC-4 BBH events, inferred using the \flowMC sampler.}\label{fig:gwtc4_reproduce_masses}
\end{figure}

\revision{
	To demonstrate the capabilities of conducting realistic population studies, we have reproduced the population analysis \cite{2025arXiv250818083T} using the 153 BBH events from the LVK's GWTC-4 population study \cite{ligo_scientific_collaboration_and_virgo_2025_17014085,ligo_scientific_collaboration_2025_16911563}. By employing the exact mass model \textsc{Broken Power Law + 2 Peaks} with smoothing at the lower mass end (Equation B18 and B21 of \cite{ligo_scientific_collaboration_2025_16911563}) and incorporating selection effects based on the injections \cite{ligo_scientific_collaboration_2025_16740128}}, we successfully recreated the population plots using \flowMC sampler in \gwk. \revision{The overplot of primary mass and mass ratio is shown in Figure~\ref{fig:gwtc4_reproduce_masses}. Posteriors used in the comparison are filtered with the variance of less than 1. See equation 9, 10 and 11 of \cite{2025PhRvD.111f3043H} for variance of the population likelihood (Equation \eqref{eq:likelihood})}.

% \begin{figure}[t!]
% 	\includegraphics[width=\columnwidth]{figures/fig_spins.pdf}
% 	\caption{\label{fig:gwtc4_reproduce_spins}\textbf{GWTC-4 Reproduction (Spins):} Spin magnitude (top) and tilt (bottom) distributions for GWTC-4 BBH events, inferred using the \flowMC sampler.}
% \end{figure}
%
% \begin{figure}[t!]
% 	\includegraphics[width=\columnwidth]{figures/fig_redshift.pdf}
% 	\caption{\label{fig:gwtc4_reproduce_redshift}\textbf{GWTC-4 Reproduction (Redshift):} Redshift evolution model for GWTC-4 BBH events, inferred using the \flowMC sampler.}
% \end{figure}

\section{Conclusion}
\label{sec:conclude}

In this paper, we presented the methodology implemented in our JAX-based framework, \gwk, designed to infer the properties of multiple populations of gravitational-wave sources. As an extension of \textsc{PopModels}, \gwk\ offers a modular and efficient inference engine accessible via a user-friendly command-line interface, enabling the construction of complex population models from simple components. It also provides functionality to generate mock catalogs, allowing researchers to explore potential astrophysical scenarios.

We demonstrated four types of analyses using \gwk. Two of them replicate previously published results: the recovery of population parameters for non-spinning eccentric binaries, and the inference of black hole mass distributions from the LIGO-VIRGO-KAGRA \revision{GWTC-4} catalog. The other two represent novel contributions. First, we showed that the eccentricity distribution of spinning eccentric binaries can be successfully recovered, demonstrating robustness in the presence of spin effects. Second, we introduced a multi-source population model with independently varying rates, illustrating \gwk's ability to disentangle multiple sub-populations simultaneously, an important step toward more realistic population synthesis studies. In contrast to previous studies that focus on mass and spin studies in sub-populations, \gwk\ provides the capability to explore additional parameters such as eccentricity and redshift distributions.

The \gwk\ framework is open-source and publicly available on GitHub \cite{git_gwkokab}, along with documentation for ease of use. The capability to model and recover properties of diverse populations is essential for advancing our understanding of compact binary formation channels. \gwk\ represents a significant step toward this goal, and we anticipate it will be a valuable resource for the gravitational-wave astrophysics community in both current analyses and preparation for future observational campaigns.

The \gwk\ framework has two principal rationales.  First and foremost, \gwk\ provides the capability to identify subpopulations whose signatures may have multiple correlated signatures (e.g., common mass features among BH-NS and BBH; correlations in mass, both component spin magnitudes and orientations; et cetera).
The model-based approach adopted by \gwk\  nominally has less flexibility than the multiple nonparametric methods which have been used to characterize the population of detected
gravitational wave transient sources, as a parameter distribution of merging
compact binaries \cite{2017MNRAS.465.3254M,2023ApJ...957...37R,2023ApJ...946...16E,2025PhRvD.111l3049M,2025PhRvD.111f3043H}. However, in practice these nonparametric methods have been applied to only a handful of dimensions, at times treating only one nonparametrically while employing strong models for others, and only encode local correlations.   For scenarios with well-motivated physical predictions across multiple observables, a model-based approach provides the sharpest conclusions.
Second, compared with other model-building frameworks, \gwk\ allows new and experienced users to quickly design, prototype and perform analyses, all essential given the rapid pace of discovery as the GW census grows.

\section*{Acknowledgements}
This material is based upon work supported by NSF's LIGO Laboratory which is a major facility fully funded by the National Science Foundation. The authors acknowledge the computational resources provided by the LIGO Laboratory's CIT cluster, which is supported by National Science Foundation Grants PHY-0757058 and PHY0823459. ROS acknowledges support from NSF Grant No. AST-1909534, NSF Grant No. PHY-2012057, and the Simons Foundation. MZ also acknowledges the support from the Fulbright program and the Higher Education Commission of Pakistan (HEC). MQ acknowledges the computational resources provided by Habib University. The authors also like to thank Asad Hussain for helping in debugging the code and useful discussions.

\section{Appendix}
\label{sec:appendix}

\subsection{Population Models}
\label{appendix:subsec:population-models}
This study adopts a form of \textsc{Powerlaw Primary Mass Ratio} where the primary mass is modeled with a power law with index \(\alpha\) and mass ratio is modeled with power law with index \(\beta\). It is detailed in Appendix B1 of \cite{2023PhRvX..13a1048A} and equivalent to the \textsc{Truncated Mass Model} detailed in Appendix B1a of \cite{2021ApJ...913L...7A}, and the \textsc{Two Sided Truncated Mass Model}, described in Section IID of \cite{2019PhRvD.100d3012W}. Similar formalism is also described in Model-C \cite{2019ApJ...882L..24A}. Evolution of redshift is modeled through \textsc{Powerlaw Redshift} \cite{2018ApJ...863L..41F, 2014ARA&A..52..415M,2023PhRvX..13a1048A}. We assume a $\Lambda$CDM cosmology using the cosmological parameters from Planck 2015 \cite{2016A&A...594A..13P}.

We implemented the truncated gaussian for spin magnitudes and also \textsc{Default Spin Model}, which was first presented by \cite{2019ApJ...882L..24A} and further developed by \cite{2019PhRvD.100d3012W}, defines how the spins of binary objects are distributed. We define \(t_i=\cos(\theta)\) for \(i=1,2\) as the cosine of the tilt angle between component spin and a binary's orbital angular momentum. We followed the assumption that \(t\) is from a mixture of an isotropic and a gaussian distribution \cite{2017PhRvD..96b3012T}. The \textsc{Half Normal Eccentricity Model} characterizes the orbital eccentricity distribution through a Half Normal probability density function \cite{2024PhRvD.110f3009Z} bounded between $0$ and $1$.
\subsection{Synthetic Population Generation}
\label{subsection:syn-pop-uncertainties}

We may want to generate synthetic population for a potential science cases. Therefore, we have provided a flexible application programming interface (API) in \gwk to generate injection for source parameter $\svec$ and add errors in them to make fake posterior estimates using a previously described schematic in section III.A of \cite{2024PhRvD.110f3009Z}. The injections are drawn in terms of primary $m_1$ and secondary $m_2$ masses, but errors are added in terms of chirp mass $\mc$ and symmetric mass ratio $\eta$ using the following relations.

\begin{figure*}
	\includegraphics[width=\textwidth]{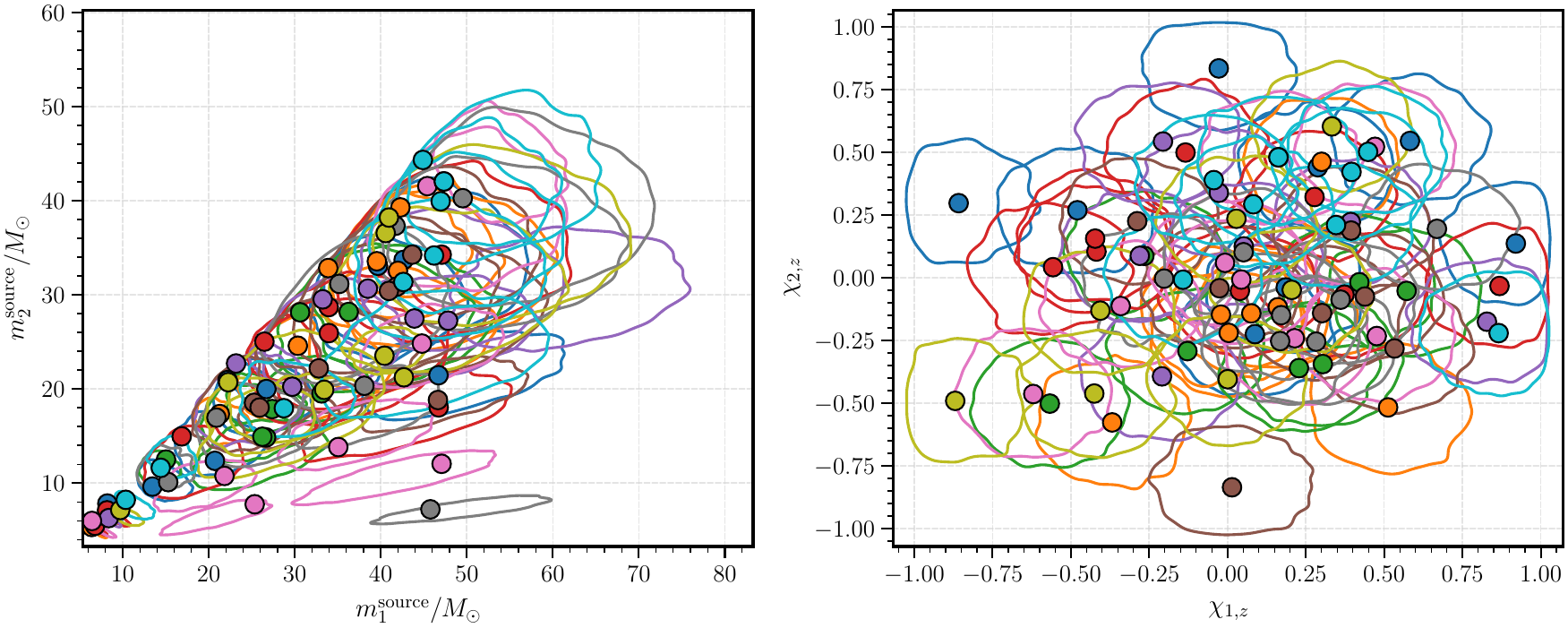}
	\caption{
		\textbf{Spinning Eccentric BBHs Synthetic Population:} The RHS shows the true injections generated using power law, and banana error being added described in Section \ref{subsection:syn-pop-uncertainties}. The LHS figure shows the true injections generated using normal distribution for spin, with error being added using the truncated gaussian distribution. The dots represent the true injections, and the contours show the error in those injections.}
		\label{fig:fake_pe}
\end{figure*}

\begin{align}
	\mc  & = \mc_{T}(1+\beta(r_0+r)),                             \\
	\eta & = \eta_{T}\left(1+0.03(r_0'+r')\frac{12}{\rho}\right).
\end{align}

Here $\mc_{T}$ and $\eta_{T}$ are converted injections from true $m_1$ and $m_2$. The $r_0$ and $r_0'$ are the random numbers drawn from the standard normal distribution, which will shift the mean of the $\mc$ and $\eta$ distribution with respect to $\mc_{T}$ and $\eta_{T}$ respectively.
The $r$ and $r'$ are the independent and identically distributed arrays of those randomly generated numbers to spread the distribution.
The measurement uncertainty is inversely proportional to signal-to-noise ratio $\rho$, drawn from the
distribution $p(\rho) \propto \rho^{-4}$, which holds for isotropically distributed sources in a static
universe, subject to the threshold $\rho\geq 8$ for detection. Following Section III.D of \cite{2022arXiv221007912W},
we estimate $\beta \simeq (6/\rho)(v/0.2)^7 $ where $v$ is an estimated post-Newtonian orbital velocity at a reference frequency of 20
Hz, and  $\rho$ is drawn from a Euclidean SNR distribution $P(>\rho)\propto 1/\rho^3$.

\gwk also generates the random injection for spin, tilt, and eccentricity using the normal distribution. If required we can also make it truncated or half normal by choosing the appropriate value of $a$, $b$, $\mu$ and $\sigma$ for lower, higher, location and scale values respectively

\begin{align}
	x_{T} = \mathcal{N}_{[a,b]}(\mu,\sigma^2).
\end{align}

\gwk also provide the option to generate redshift samples consistent with the assumed redshift evolution model,

\begin{equation}\label{eq:Z_data}
	q(z|\zvec_i)\propto\frac{1}{1+z}\frac{dV_c}{dz}(1+z)^{\zvec_i},
\end{equation}

we normalize this function over the redshift range of interest $[0,z_{\mathrm{max}}]$, to obtain the normalization constant $\mathcal{Z}_i$ for each population $i$, and it is given by,

\begin{equation}\label{eq:Z*_i}
	\mathcal{Z}_i(\kappa_i,z_{\mathrm{max}})=\int_0^{z_{\mathrm{max}}} \frac{1}{1+z}\frac{dV_c}{dz}(1+z)^{\zvec_i} dz.
\end{equation}

Thus we will get the normalized redshift distribution for each population $i$ as follows,

\begin{equation}\label{eq:redshift_distribution}
	p(z|\zvec_i)=\frac{1}{\mathcal{Z}_i(\kappa_i,z_{\mathrm{max}})}\frac{1}{1+z}\frac{dV_c}{dz}(1+z)^{\zvec_i}.
\end{equation}

After generating the injections (true values) for $(\svec,z)$, we can use an independent gaussian with the $\mu=x^T$ true value as a location and flexible value of $a$, $b$, and $\sigma$ to add the desired errors in the synthetic population. The final posterior distribution for spin, tilt, and eccentricity is given by

\begin{align}
	x = \mathcal{N}_{[a,b]}(x_{T},\sigma^2).
\end{align}

The final injections and posteriors for spinning eccentric population are shown in Figure~\ref{fig:fake_pe} which show the shape of the error introduced in the synthetic population. The true injections are shown in the dots and the contours show the error in those injections.

\subsection{Hierarchical Bayesian Modeling}
\label{subsec:appendix:hbm}

%Lets start with models with redshift evolution.
The likelihood of individual events using Bayes theorem is given by,
\begin{equation}
	\label{eq:single_event_likelihood}
	\ell(\svecz)=p(d|\svecz)\propto \frac{p(\svecz|d)}{\pi(\svecz)},
\end{equation}
where $\pi(\svecz)$ is the reference prior probability of the binary parameters (intrinsic and extrinsic).
The next is population likelihood which is the most crucial part of population inference given in Equation \eqref{eq:likelihood} with independent rate and redshift evolution for each population. It can be expanded as follows,

\begin{multline}\label{eq:likelihood_expanded}
	\int \ell_j(\svecz) \comp(\svecz\mid\pvecz)   \sqrt{ g_{\svecz}}d\svecz \propto \sum_{i=1}^M\iint \frac{p(\svecz|d_j)}{\pi(\svecz)} \\\srate_{0_i}  p_i(\svec|\pvec_i)  (1+z)^{\zvec_i}  T_{\mathrm{obs}} (1+z)^{-1} (dV_c/dz) d\svec dz.
\end{multline}

We used importance sampling to estimate the integral in the likelihood function and can be approximated for a $j^{th}$ event as follows:
\begin{equation}\label{eq:likelihood_estimate}
	\left\langle\frac{\comp(\svecz_{j,k}\mid\pvecz) T_{\mathrm{obs}} (1+z)^{-1} (dV_c/dz)}{\pi(\svecz_{j,k})}\right\rangle _{\svecz_{j,k}\sim p(\svecz| d_j)},
\end{equation}

where $\svecz_{j,k}$ can be provided through data files of individual events and $k$ is the number of posterior samples in each event. Along with cosmo PE samples from data files, we import the following reference prior from \textsc{Bilby} \cite{bilby_paper} with normalization constant $V_0=\int_0^{z_{\mathrm{max}}}{\frac{dV_c}{dz}\frac{1}{1+z}}dz$. Our reference prior is based on the assumption that the intrinsic parameters of the binary system are uniformly distributed in the source frame and can be expressed as,

\begin{equation}\label{eq:reference_prior}
	\pi(\svecz) = \frac{1}{V_0}\underbrace{\frac{dV_c}{dz}\frac{1}{1+z}}_{\text{Uniform in Source Frame}}\times \underbrace{(1+z)^2}_{\text{Masses in Source Frame}}.
\end{equation}

However, appendix C of GWTC-2 \cite{2019ApJ...882L..24A} have defined alternative reference priors along with non-cosmo PE samples,

\begin{equation}
	\pi(\svecz) = 4 \pi ~~ d^2_L(z)\frac{\partial d_L(z)}{\partial z}\times \underbrace{(1+z)^2}_{\text{Masses in Source Frame}}
\end{equation}
where \(d_L(z)\) is the luminosity distance in the source frame.

When working with synthetic (fake) data, we typically adopt constant reference prior. However for real data analysis, the choice of reference prior requires careful consideration, as an inappropriate prior can introduce significant biases in the population inference \cite{2022RvMP...94b5001C,ThraneTalbot2019,2014PhRvD..89f4048O,2023PhRvX..13d1039A}. Historically, parameter estimation (PE) for gravitational-wave events has used priors that are uniform in detector-frame masses. In contrast, population inference requires astrophysically motivated priors, typically uniform in source-frame masses. To reconcile this difference, we convert the PE prior into the source frame using the Jacobian transformation $dm_{\svec,\mathrm{source}}=dm_{\svec,\mathrm{det}}/(1+z)$ applied to each component of the parameter vector $\svec$. The analytic form of the PE priors and their parameter ranges are usually documented in the data release metadata. The most recent PE data releases \cite{2025arXiv250818081T, ligo_scientific_collaboration_and_virgo_2025_17014085} also provide cosmologically reweighted posterior samples, which can be directly used in population studies when combined with appropriately converted source-frame reference priors.

To prevent floating point numbers from underflow, the optimization is
applied to the logarithm of the likelihood function. Logarithms are
monotonically increasing functions and their composition preserves
relative minima and maxima. The logarithm of the likelihood function is given by,
\begin{multline}
	\ln\mathcal{L}(\pvecz)
	\propto
	-\mu(\pvecz)+
	\sum_{j=1}^N\\
	\ln\left( \left\langle\frac{\comp(\svecz_{j,k}\mid\pvecz) T_{\mathrm{obs}} (1+z)^{-1} (dV_c/dz)}{\pi(\svecz_{j,k})}\right\rangle _{\svecz_{j,k}\sim p(\svecz| d_j)}\right).
\end{multline}

Similarly, we can expand Equation \eqref{eq:mu} with Equation \eqref{eq:total_rate} to get the expected number of detections,
%
% \begin{align}
%     \label{eq:mu_expanded}
%     \mu(\pvec,\zvec)= T \iint \frac{dV_c}{dz}\frac{P_{\mathrm{det}}(\svec;z)}{1+z}\sum_{i=1}^{M}\comp_i(\svec,z\mid\pvec_i,\zvec_i)
%     dz d\svec.
% \end{align}
%
This will provide us normalized probability distribution of redshift for each population $i$,
\begin{equation}\label{eq:mu_expanded}
	\hat{\mu}(\pvecz)=\sum_{i=1}^{M}T_{\mathrm{obs}} \mathcal{Z}_i\int P_{\mathrm{det}}(\svec;z) \srate^*_i(\pvec_i) p_i(\svecz|\pvecz_i)d\svecz,
\end{equation}
where $p_i(\svecz|\pvecz_i) = p_i(\svec|\pvec)\times p_i(z|\zvec)$ is a normalized probability distribution of intrinsic and extrinsic parameter $\pvecz$. It allows us to estimate the expected number of detections by drawing samples from $p_i$ and using importance sampling to estimate the integral as follows,
\begin{equation}\label{eq:mu_expanded2}
	\hat{\mu}(\pvecz)=\sum_{i=1}^{M}T_{\mathrm{obs}}\mathcal{Z}_i \srate^*_i(\pvec_i)\left\langle P_{\mathrm{det}}(\svec_{p,i};z_{p,i})
	\right\rangle _{\svecz_{p,i}\sim p_i(\svecz\mid\pvecz_i)}.
\end{equation}

In case of incorporating the realistic sensitivity effects, we use injections available on \cite{ligo_scientific_collaboration_2025_16740128}. We use Equation A2 of \cite{2023PhRvX..13a1048A} to compute the estimated rate. Our implementation can be shown as,

\begin{equation}\label{eq:realistic_mu}
	\hat{\mu}(\pvecz) \approx \frac{1}{N_{\mathrm{total}}}\sum_{i=1}^{N_{\mathrm{found}}}\frac{\comp(\svecz_{i}\mid\pvecz)T_{\mathrm{obs}} (1+z)^{-1} (dV_c/dz)}{\pi_{\mathrm{draw}}(\svecz_{i})}
\end{equation}

where \(N_{\mathrm{total}}\) is the total number of injections, \(N_{\mathrm{found}}\) is the number of injections that are found in the data, \(\pi_{\mathrm{draw}}(\svecz_{i})\) is the drawing probability or sampling pdf of the injection.
The final step is to compute the posterior distribution of the population parameters $\pvec$ by taking the product of the likelihood function and the prior distribution of the population parameters as given in Equation \eqref{eq:Bayes_pop}.
% The complete flow of the likelihood function is shown in Figure \ref{fig:likelihood}. In addition to that, the API structure is to make inference without playing with python file is also shown in Figure \ref{fig:sage_pipeline}. 
The API is designed to be flexible and user-friendly, allowing users to easily customize their analysis without needing to modify the underlying code.

\subsubsection{Non-Evolving Redshift Models}

For the non-evolving redshift models, we can use the same Equation \eqref{eq:likelihood} by putting $\zvec=0$ which will remove the redshift evolution factor \( (1+z)^{\zvec} \). We follow the same procedure as described above to estimate the expected number of detections and posterior distribution of the population parameters. The only difference is that we do not need to sample from the redshift distribution and likelihood function will be simplified to,

\begin{equation}\label{eq:non_evolving_likelihood}
	\mathcal{L}(\pvec,0) \propto
	e^{-\mu{(\pvec,0)}}
	\prod_{j=1}^N
	\iint\ell_j(\svec,z) \comp(\svec,z\mid\pvec,\zvec=0)
	d\svec dz.
\end{equation}

and similarly the expected number of detections $\hat{\mu}(\pvec,0)$ will be given by,

\begin{equation}\label{eq:mu_expanded_non_evolving}
	\sum_{i=1}^{M}T\mathcal{Z}_i \srate^*_i(\pvec_i)\left\langle P_{\mathrm{det}}(\svec_{p,i};z_{p,i})
	\right\rangle _{\svec_{p,i},z_{p,i}\sim p^{*}_i(\svec,z\mid\pvec_i,\zvec_i=0)}.
\end{equation}

% Importantly, if $P_{\mathrm{det}}(\svec;z)$ lacks the redshift dependence (computed on fixed redshift), the recovery of the redshift evolution parameter $\zvec_i$ may be baised. To avoid this biasness under the assumption of $\zvec_i=0$, it is essential to ensure that $P_{\mathrm{det}}(\svec;z)$ is independent of redshift. Conversely, to recover the redshift evolution parameter $\zvec_i$, it is necessary to include the redshift evolution in $P_{\mathrm{det}}(\svec;z)$ which is more physically motivated approach.

% In the case of redshift-independent models, the simplified expected rate of detections from equation \eqref{eq:mu_expanded} can be approximated by simple Monte Carlo integration,

% \begin{align}
%     \mu(\pvec) = \srate^*\left<V(\svec)\right>.
% \end{align}

% where \(V\) is the expected comoving volume of the population model defined as

% \begin{align}
%     V(\svec) = \int \frac{dV_c}{dz}\frac{P_{\mathrm{det}}(\svec)}{1+z} dz,
% \end{align}

% \begin{figure}[t]
%     \includegraphics[width=\columnwidth]{figures/sage_pipeline.pdf}
%     \caption{\label{fig:sage_pipeline} \textbf{Inference using sage:} This pipeline shows the complete flow of sage in \gwk. Each command line in \gwk begins with sage follows this structure to run the population inference.}
% \end{figure}

% \begin{figure*}[t]
%     \includegraphics[width=\textwidth]{figures/likelihood.pdf}
%     \caption{\label{fig:likelihood} \textbf{Posterior Probability:} This figure shows the complete calculation structure of posterior probability in \gwk.}
% \end{figure*}

\subsection{Posterior Predictive Distribution}
\label{subsec:ppd}
The PPD is the distribution of future data given the observed data. It is given by,

\begin{equation}\label{eq:ppd}
	p(d_{\text{future}}| d_{\text{obs}}) = \underset{\pvec\sim p(\pvec| d_{\text{obs}})}{\mathbb{E}}\left[p(d_{\text{future}}|\pvec)\right],
\end{equation}

where \(d_{\text{future}}\) is the future data, \(d_{\text{obs}}\)
is the observed data, and \(\pvec\) is the model parameters. It can
be approximated by the Monte Carlo method by drawing sufficiently
large (\(N\)) samples (\(\pvec_i\)) from the posterior
distribution. The PPD is then approximated by,

\begin{equation}\label{eq:ppd_approx}
	p(d_{\text{future}}| d_{\text{obs}}) \approx \frac{1}{N}\sum_{i=1}^{N}p(d_{\text{future}}|\pvec_i),
\end{equation}

% Explanation: We have removed the ppd CLIs
% We have also made the command line to generate the PPD plots with confidence intervals. Confidence intervals are made by calculating each quantile of the distributions. Additionally, users can specify the desired confidence level for the intervals to customize their analysis.

%\nolinenumbers

\bibstyle{apsrev4-2}
\bibliography{references.bib}

\end{document}